\documentclass[12pt]{article}
\usepackage[utf8]{inputenc}
\usepackage{graphicx}
\usepackage{authblk}
\usepackage[section]{placeins}
\usepackage{amsmath}
\usepackage[top=2.0cm,bottom=2.0cm,left=3.0cm,right=1.5cm]{geometry}
\usepackage{float}
\usepackage{setspace}   
\usepackage{amsfonts} 
\usepackage{caption}
\usepackage{subcaption}
\usepackage{hyperref}
\hypersetup{
	citecolor = blue,
	colorlinks = true,
	linkcolor = blue,
	filecolor = magenta,      
	urlcolor = cyan,
}
\usepackage[normalem]{ulem}
\usepackage[table,xcdraw]{xcolor}
\usepackage{multirow}

\newcommand\flast{\bgroup\markoverwith{\textcolor{orange}{\rule[0.5ex]{1pt}{1pt}}}\ULon} 

\title{Assortativity in sympatric speciation and species classification}

\author[1]{Joao U. F. Lizárraga}
\author[1,2]{Flavia M. D. Marquitti\thanks{Corresponding author: flaviam@unicamp.br}}
\author[1]{Marcus A. M. de Aguiar}
\affil[1]{Instituto de F\'{i}sica Gleb Wataghin, Universidade Estadual de Campinas, Campinas, Brazil}
\affil[2]{Instituto de Biologia, Universidade Estadual de Campinas, Campinas, Brazil}
\date{}
\begin{document}
	
\maketitle
	
\section*{Abstract}

We investigate the role of assortative mating in speciation using the sympatric model of Derrida and Higgs. The model explores the idea that genetic differences create incompatibilities between individuals, preventing mating if the number of such differences is too large. Speciation, however, only happens in this mating system if the number of genes is large. Here we show that speciation with small genome sizes can occur if assortative mating is introduced. In our model individuals are represented by three chromosomes: one responsible for reproductive compatibility, one for coding the trait on which assortativity will operate, and a neutral chromosome. Reproduction is possible if individuals are genetically similar with respect to the first chromosome, but among these compatible mating partners, the one with the most similar trait coded by the second chromosome is selected. We show that this type of assortativity facilitates speciation, which can happen with a small number of genes in the first chromosome. Species, classified according to reproductive isolation, dictated by the first chromosome, can display different traits values, as measured by the second and the third chromosomes. Therefore, species can also be identified based on similarity of the neutral trait, which works as a proxy for reproductive isolation.

\noindent{\textbf{Keywords:}} assortative mating, homotypic preferences, speciation, two-allele model
	
\section*{Introduction}

Speciation results from the interplay of different isolating mechanisms \cite{coyne2004speciation,nosil2012ecological}, such as geographic isolation \cite{mayr2013animal,Manzo_Peliti_1994,Yamaguchi-2013_first,princepe2022diversity}, genetic incompatibilities \cite{nosil2011genes}, competition for resources \cite{Dieckmann_1999,Dieckmann_2003,polechova2005speciation}, and temporal separation \cite{cooley2003temporal,taylor2017role}. Ultimately, speciation is the result of significant decrease in gene flow between groups of individuals, allowing adaptations and random changes occurring in each group to be restricted to that group. When this happens, genetic and phenotypic differences can accumulate over time, leading to   reproductive isolation and possibly to hybrid incompatibility. Assortative mating has been conjectured to reinforce the speciation process by preventing individuals with different phenotypes to mate when they come into contact. Assortative mating occurs when individuals with similar phenotypes mate more often than would be expected at random. Frequent mating of dissimilar individuals would mix their genomes, eventually leading to the reversion of  speciation.  Evidence demonstrating the role of assortative mating in different cases has accumulated \cite{mckinnon2004evidence,janicke2019assortative, kopp2018mechanisms}, although it might be inneffective in hybrid zones \cite{irwin2020assortative}. Theoretical studies have contributed to these claims as well ~\cite{bagnoli2005model, schneider2014toward, schneider2015dynamical}. 

In this work we investigate the role of assortativity in a simple model of speciation. For that, we focus on sympatric speciation, i.e., the development of reproductive isolation without geographic barriers.
In this case, mating is not restricted by the presence of physical barriers or by spatial proximity of individuals, but it cannot be completely random either, otherwise no reduction in gene flow can occur. Sympatric speciation, therefore, requires some form mating selection \cite{nosil2012ecological}, such as that promoted by temporal isolation \cite{boumans2017ecological}, genetic incompatibilities \cite{rieseberg2010speciation} or assortative mating based on phenotypic characters \cite{kondrashov1998origin,kopp2018mechanisms,doebeli2000evolutionary,caetano2020sympatric}. Empirical evidence for sympatric speciation is reviewed in \cite{Bolnick_2007}. Here we explore a model of sympatric speciation driven be genetic incompatibilities and assortative mating.

From a theoretical point of view two models of sympatric speciation have stood out. The first, proposed by Dieckmann \& Doebeli \cite{Dieckmann_1999} argues that strong competition for resources could drive speciation even without any form of geographic isolation. If resources are characterized by a continuous parameter, such as seed size, the theory suggests that it might be more advantageous for individuals who have phenotypes adapted to consume extreme resources (such as very small or very large seeds, of which there are fewer) than it is for those with a more common phenotype -- for which there are abundant resources but also strong competition -- resulting in a disruptive selection. In this case two stable groups would emerge, adapted to the extremes of the resource distribution, whereas   intermediate phenotypes would have lower fitness. The model has been criticized for requiring unrealistic high mutation rates \cite{polechova2005speciation}.

The second theoretical model, proposed by Higgs \& Derrida \cite{higgs1991stochastic,higgs1992genetic, gavrilets1998rapid,de2009global} (DH model), demonstrated that, even without competition, sympatric speciation may still occur if mating is restricted by genetic similarity. The model is based on the idea that genetic differences create incompatibilities between individuals and, if the number of such differences is too large, mating would not happen. The model considers hermaphrodite individuals whose genomes are described by binary chains of biallelic genes. Mating occurs only between individuals whose genetic similarity is large enough -- i.e. based on  prezygotic barriers. Speciation, however, only happens if the number of genes responsible for creating these incompatibilities is sufficiently large, of the order of thousands of genes, depending on the model parameters \cite{de2017speciation}. The original model considered only infinitely large genomes to avoid this limitation.

The need for very large genomes in the DH model can be circumvented by adding auxiliary mechanisms that contribute to further reduce gene flow among the individuals. One possibility is to introduce space and restrict mating not only by genetic similarity but also by spatial proximity \cite{de2017speciation, de2009global}. This facilitates speciation and drastically reduces the number of genes required for the formation of reproductively isolated groups. Speciation, however, is now {\it parapatric}, as there is isolation by distance, a weak form of geographic isolation. 

Here we reexamine the DH model with finite genomes and consider the effects of assortative mating in facilitating speciation. Assortative mating has been observed in many species,especially with respect to body size, pheromones and coloration \cite{rosenthal2017mate}, and it can be a powerful driver of speciation. For instance, Puebla et al 2012 \cite{puebla2012pairing} found that the pairing dynamics of hermaphrodite fish of the genus \textit{Hypoplectrus} is related to color patterns that is used as an indicator to release sperm and eggs into the water. In order to introduce assortative mating in the DH model, we split the genomes into three independent chromosomes. This division in chromosomes serves solely to identify genes that have similar roles. They differ from real chromosomes in the sense they are not physically linked during reproduction. We then have, one responsible for creating reproductive incompatibilities as in the original model, one for coding the trait on which assortativity will operate, and one neutral chromosome coding for a second trait which is not under selection by assortativity and is not responsible for reproductive incompatibilities. In this way, reproduction is possible only if individuals are genetically similar with respect to the first chromosome, but among these compatible mating partners, the one with the most similar assortativity chromosome will be chosen. We show that: (i) assortativity has a dramatic effect on speciation, greatly facilitating the process; (ii) species, classified according to reproductive isolation dictated by the first chromosome, can display different traits values, as measured by the second and the third chromosomes; (iii) this implies that species identification based on similarity of the assortativity trait or the neutral trait generally coincides with that based on reproductive isolation.

We simulated the evolution of a finite population with sexual reproduction and mutation based on the Derrida-Higgs (DH) algorithm \cite{higgs1991stochastic, higgs1992genetic} -- an individual based model (IBM). We modified the original model to contemplate the possibility of choosiness via assortative mating and introduced a neutral chromosome as an independent marker of the genetic evolution. We then analysed the effects of different levels of assortativity in the speciation process. Below we review the three original models, describe our adaptations, and summarize the parameters used in our simulations.

\subsection*{The DH models}

Derrida and Higgs proposed three models to describe the evolution of populations of size $M$ with different forms of reproduction: (i) asexual; (ii) sexual with random mating; (iii) sexual with mating constrained by genetic similarity \cite{higgs1991stochastic,higgs1992genetic}. In all cases the individuals were considered hermaphrodite and haploid, with genomes represented by a single chromosome $\mathbb{S}$: 

\begin{equation}
	\mathbb{S}^{\alpha}_{F}:\{S_{1}^{\alpha}, S_{2}^{\alpha}, S_{3}^{\alpha}, \ldots, S_{F}^{\alpha}\}.
	\label{eq:genome}
\end{equation}
Here $\alpha$ labels the individual, $F$ is the number of \textit{loci} and $S^{\alpha}_i$ represents a biallelic gene at locus $i$, which can take the values $\pm 1$. Gene and allele are interchangeable nomenclatures here because we are considering each \textit{locus} as a gene and because individuals are haploid, the allele of a given gene is representing the gene itself. In the DH models the authors consider the limit $F \rightarrow \infty$, as this simplifies the simulations and the dynamics, allowing the derivation of several analytical results.

The population is characterized by a matrix $d_{F}^{\alpha\beta}$ containing the normalized genetic distance between all pairs of individuals $\alpha$ and $\beta$:
\begin{equation}
	d_{F}^{\alpha\beta} = \frac{1}{F}\sum_{i = 1}^{F}{|S_{i}^{\alpha} - S_{i}^{\beta}|}.
	\label{eq:hamming}
\end{equation}
The population can also be characterized by the genetic similarity
\cite{higgs1992genetic}
\begin{equation}
	q_{F}^{\alpha\beta} = \frac{1}{F}\sum_{i = 1}^{F} S_{i}^{\alpha} S_{i}^{\beta}
	\label{eq:sim}
\end{equation}
which is related to the genetic distance by
\begin{equation}
	q_{F}^{\alpha\beta} = 1 - d_{F}^{\alpha\beta}
	\label{eq:rel}
\end{equation}
with $0 < d_{F}^{\alpha\beta} < 2$ and $-1 < q_{F}^{\alpha\beta} < 1$. 

Initially, all individuals have identical genomes, with $S_i^\alpha=1$. The evolution of a population with $M$ individuals is performed as follows:

{\bf (i) asexual model}: an individual is randomly chosen from the population to reproduce; an offspring is created, receiving an exact copy of its parent's genome; 

{\bf (ii) sexual model with random mating}: two individuals are randomly chosen from the population to reproduce; an offspring is created by recombining the genomes of the parents, such that the allele for each gene is received from either parent with equal probability; 

{\bf (iii) sexual model with mating restriction}: a first parent $\alpha$ is randomly chosen from the population, but the second parent $\beta$ is chosen only from those individuals having $d_{F}^{\alpha\beta} \leq g$. If no such individual can be found, the first parent is discarded and a new one is chosen, repeating the search for a compatible mate until an offspring is produced. The parameter $0<g<1$ sets the maximum genetic distance for reproductive compatibility.


In all cases the individuals are selected with replacement, so that the same individual can be selected more than once. Also, after the offspring has been created, each of its genes is allowed to mutate with rate $\mu$ and the whole process is repeated until $M$ offspring are created, forming the next generation.  

The evolution of genetic distances between individuals can be computed considering the relationship between parents and their offspring. Consider the case of sexual reproduction: suppose $P_{1}(\alpha)$ and $P_{2}(\alpha)$ are the parents of individual $\alpha$ and $P_{1}(\beta)$ and $P_{2}(\beta)$ are the parents of $\beta$. If $\tilde\mu$ is the mutation rate, the probability of mutation in a unit time step (one generation) is $\mu = (1-e^{-2\tilde\mu})/2$ and the probability that a mutation does not happen is $(1+e^{-2\tilde\mu})/2$. If gene $S_i^\alpha$ is inherited from $P_{1}(\alpha)$, then: 

\begin{align*}
	\mathbb{P}(S_{i}^{\alpha} &= S_{i}^{P_1(\alpha)}) = \frac{1}{2}(1+e^{-2\tilde\mu}),\\
	\mathbb{P}(S_{i}^{\alpha} &= -S_{i}^{P_1(\alpha)}) = \frac{1}{2}(1-e^{-2\tilde\mu}), 
\end{align*}
are the probabilities of keeping the allele of $P_{1}(\alpha)$ or mutating, respectively. The expected value is $E(S_{i}^\alpha) = e^{-2\tilde\mu}S_{i}^{P_1(\alpha)}$. Since the gene is inherited from either parent with equal probability, $E(S_{i}^\alpha) = e^{-2\tilde\mu}(S_{i}^{P_1(\alpha)}+S_{i}^{P_2(\alpha)})/2$. From Eq. (\ref{eq:sim}) we find, for independent genes, 
\begin{equation}
	E(q^{\alpha\beta}) = \frac{e^{-4\tilde\mu}}{4}(q^{P_{1}(\alpha)P_{1}(\beta)} + q^{P_{2}(\alpha)P_{1}(\beta)} + q^{P_{2}(\alpha)P_{1}(\beta)} + q^{P_{2}(\alpha)P_{2}(\beta)}).
	\label{eq:exp1}
\end{equation}
In the limit $F \rightarrow \infty$ the expected similarity value coincides with the realized one. For finite genomes this is only an approximation. In terms of genetic distances, using Eq. (\ref{eq:rel}), we find
\begin{equation}
	d^{\alpha\beta} = 1-e^{-4\tilde\mu}+\frac{e^{-4\tilde\mu}}{4}(d^{P_{1}(\alpha)P_{1}(\beta)} + d^{P_{2}(\alpha)P_{1}(\beta)} + d^{P_{2}(\alpha)P_{1}(\beta)} + d^{P_{2}(\alpha)P_{2}(\beta)})
	\label{eq:exp}
\end{equation}
where we have dropped the expectation symbol.
In the asexual case, since the offspring comes from a single parent, we obtain $E(q^{\alpha\beta}) = e^{-4\tilde\mu} q^{P(\alpha) P(\beta)}$.

\begin{figure}
	\centering
	\includegraphics[width=\textwidth]{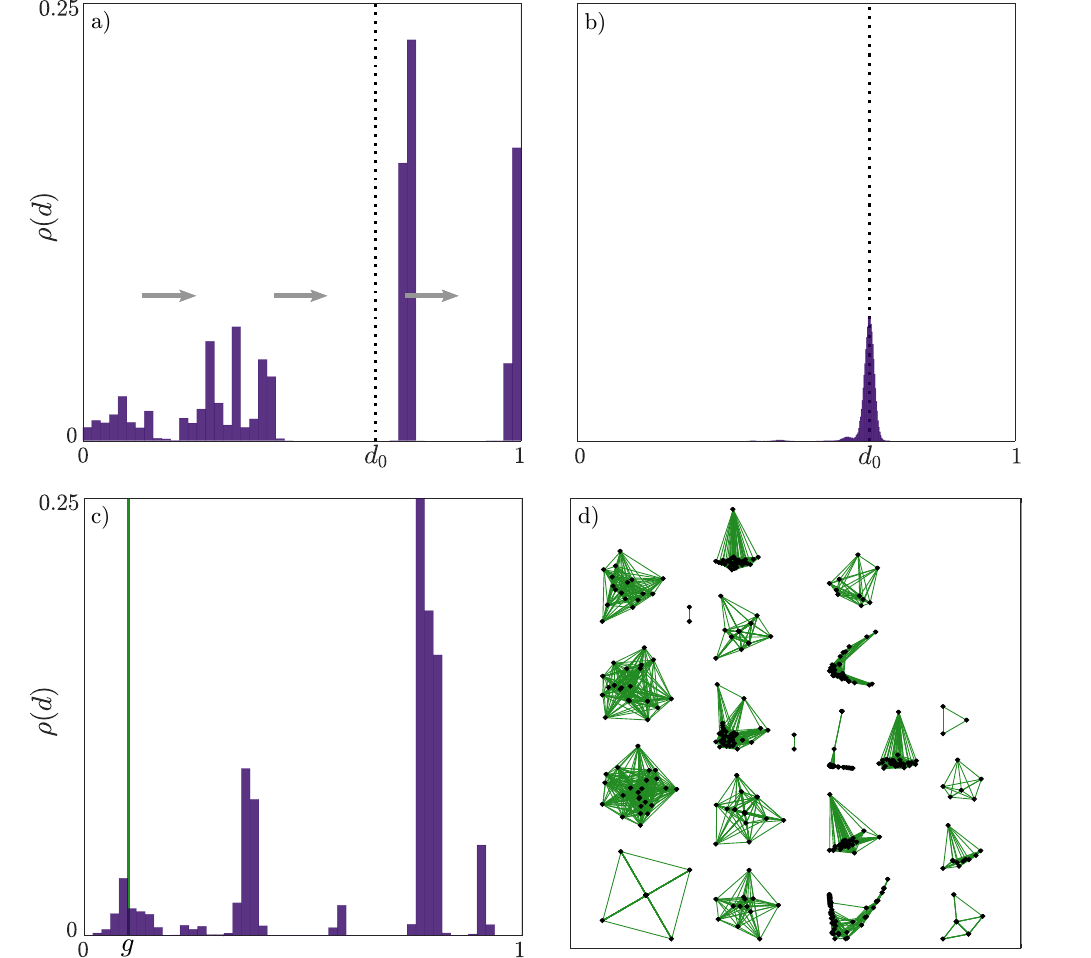}
	\caption{Genetic distance distributions in the (a) asexual, (b) sexual with random mating, and (c) sexual with mating restriction cases of the DH models at $t = 1000$. Simulations were performed for $M = 500$, $F = 10,000$, and $\mu = 10^{-3}$. In (a), arrows represent drifting of the distribution towards $1$. Dotted lines are positioned at the expected mean distance $d_0$. The distance threshold in the species-formation model was set at $g = 0.05$. Panel (d) show species representation in terms of components of the compatibility network, corresponding to the distribution shown in (c).} 
	\label{fig:fig01}
\end{figure}

In \cite{higgs1991stochastic} it is shown that the average genetic distance converges to the asymptotic value 
\begin{equation}d_{0} \approx \frac{1}{1 + (4M\tilde\mu)^{-1}},
	\label{eq:d0_ref1}
\end{equation}
where the approximation holds for large $M$ and small $\tilde\mu$. Although this result works for the cases of asexual and sexual with random mating, the distribution of genetic distances is very different in each case, as illustrated in Fig. \ref{fig:fig01} for $M=500$ individuals, $F=10,000$ loci and mutation probability $\mu=10^{-3}$. In all cases the distribution starts as a single peak at $d=0$ in $t=0$. For $t>0$ the peak broadens and moves towards $d=1$ as genomes accumulate mutations. For asexual reproduction, Fig. \ref{fig:fig01}(a), the distribution breaks into several peaks representing clusters of individuals, or strains, whose genetic distance is small within the cluster (peaks close to $d=0$) and large between clusters (peaks close to $d=1$). The distribution never reaches an equilibrium, but the average of $d$ over many realizations of the dynamics is still given by $d_0$ \cite{higgs1991stochastic}. For the random mating sexual model, on the other hand, the distribution converges to a single peak around $d_0$, showing that the population does not break into clusters, suggesting the existence of a single species. Fig. \ref{fig:fig01}(c) shows the distribution of genetic distances for the sexual model with mating restriction for $g=0.05$. In this case the distribution does break into peaks, similar to the asexual model. The peaks always move towards $d=1$ and, as they drift, many disappear, whereas new small peaks are constantly being created near d=0 \cite{higgs1991stochastic}. The result is a dynamical process of speciation and extinctions that never reaches an equilibrium. In order to understand the population structure, it is important to introduce an appropriate concept of species for these models.\\

{\bf Species definition:} in the context of the models discussed in this work, species are defined by reproductive mating affinity and are best visualized in terms of the {\it compatibility network}, which can be understood as the possible genetic flow network, as shown in Fig. \ref{fig:fig01}(d). In the network, nodes represent individuals and links are drawn between sexually compatible individuals, satisfying $d \leq g$. Species are then the components of the resulting network. Notice that components are not necessarily fully connected, implying that within species there might be sexually incompatible individuals that are indirectly connected by other members of the species via gene flow.\\

The peaks in the distribution of genetic distances in Fig.  \ref{fig:fig01}(c) are, therefore, the result of speciation. The species at time t=1000 are shown in Fig.\ref{fig:fig01}(d) for the sexual model with mating constraint. In this case, as time passes, the distribution moves to the right and reaches $d=g$, where the population breaks into species because individuals with $d>g$ cannot mate. Pairwise distances between compatible individuals (from the same species) are represented by peaks in the region $d < g$. Because species may have a few incompatible individuals that are linked indirectly via other individuals of the same species, these peaks may have a tail into $d>g$. Peaks centered at $d > g$, on the other hand, represent genetic distances between individuals of different species and, therefore, are a signature of speciation.


Two important features of the sexual model with mating restriction are: (i) in the limit $F \rightarrow \infty$ the condition $g < d_0$, with $d_0$ calculated as in Eq.~\eqref{eq:d0_ref1}, is necessary and sufficient for speciation; (ii) for finite $F$ there is a threshold $F_c$ below which speciation does not occur even if $g < d_0$. For $F < F_c$ the distribution of genetic distances behaves similarly to the random mating model, but it equilibrates at $d=g$ instead of $d=d_0$ \cite{de2017speciation}. For the parameters of Fig. \ref{fig:fig01} we find $F_c \approx 4,000$ loci.


\subsection*{The Three Chromosome Assortativity model (3CADH) }

To generalize the genetic structure of the individuals and include assortative mating, we extend the genetic model from one to three chromosomes. Each chromosome  plays a different role which we named compatibility, assortativity, and neutral, composed by $C$, $A$, and, $N$ \textit{loci}, respectively. We call them chromosomes because of their role, but they could be represented in as a single set of genes since all of them are independently passed to the offsprings. Each of these chromosomes can be defined as in \eqref{eq:genome}, changing the length $F$ to the respective length $C$, $A$, or $N$. Then, we can define the full genome of an individual $\alpha$, composed by $F=C+A+N$ loci, as the concatenation
\begin{equation}
	\mathbb{S}^{F, \alpha}: \{\mathbb{S}^{A, \alpha}, \mathbb{S}^{N, \alpha}, \mathbb{S}^{C, \alpha}\}.
\end{equation}
We name the model as ``three chromosome assortativity Derrida-Higgs'' model, or  3CADH for short.
The DH models are recovered from the 3CADH by making $A=N=0$. The genetic distance between two individuals can also be calculated with respect to each chromosome separately, adapting Eq. \eqref{eq:hamming} to the respective chromosome length. 

Evolution with assortativity is implemented in the sexual version of the model as follows: 
\begin{enumerate}
	
	\item a first parent $\alpha$ is drawn at random for reproduction; 
	
	\item the subset of individuals ${\cal C}$ compatible with $\alpha$ is selected, considering only the genetic distance between the compatibility chromosomes, i.e., satisfying the condition $d_{C}^{\alpha\beta}\leq g_{C}$; 
	
	\item from the set ${\cal C}$, potential mating partners are selected by genetic similarity in the assortative chromosome. They must satisfy the condition $d_{A}^{\alpha\gamma} \leq \delta^\alpha_A +r$ where $\delta^\alpha_A$ is the minimum distance $d_{A}^{\alpha\gamma}$ between $\alpha$ and all individuals $\gamma \in {\cal C}$ and $r$ is the choosiness parameter, measured as a fraction of $A$. Among these individuals one is randomly selected for mating. Strict assortativity corresponds to $r=0$ (mating with the most similar) whereas less stringent assortativity is obtained with $r>0$.
	
\end{enumerate}

\begin{figure}
	\centering
	\includegraphics[width = \textwidth]{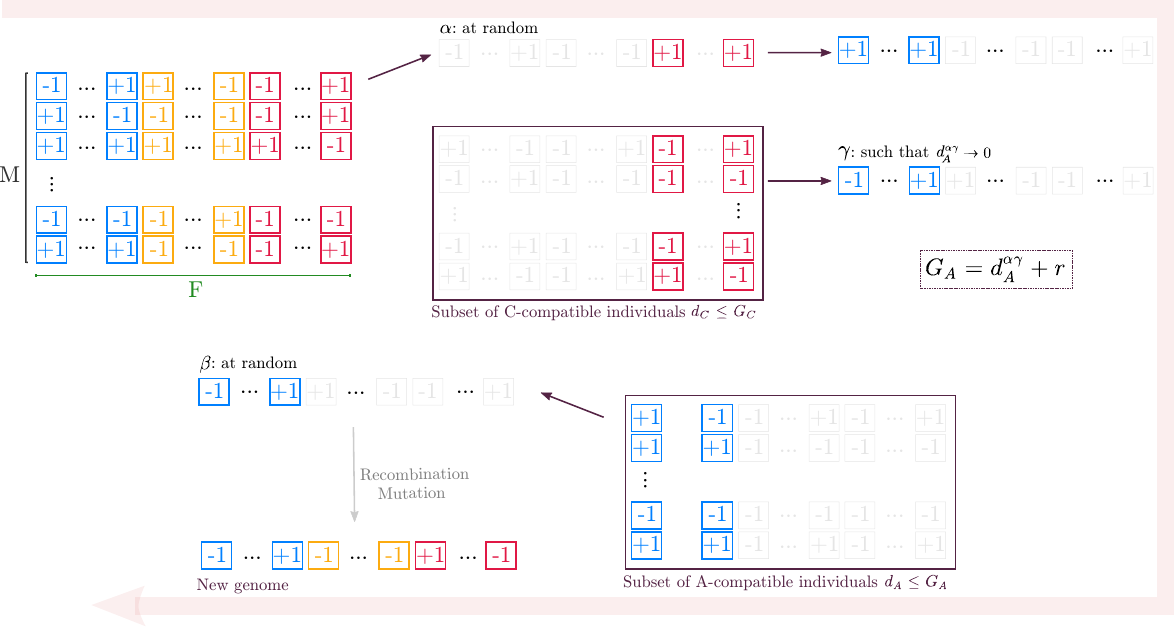}
	\caption{Population evolution in the 3CADH model. From the population at generation $t$ (top-left), an individual $\alpha$ is chosen at random for reproduction. Individuals sexually compatible with $\alpha$ are then arranged in a subset ${\cal C}$ (top-middle). We compute $\delta^\alpha_A$ as the minimum distance $d_{A}^{\alpha\gamma}$, with respect to the assortativity chromosome, between $\alpha$ and all individuals $\gamma \in {\cal C}$. The potential mates of $\alpha$ are contained in another subset ${\cal M}$ such that $d_{A}^{\alpha\gamma} \leq \delta^\alpha_A + r$ where $r$ is the choosiness parameter. Strict assortativity corresponds to $r=0$. A mating partner $\beta$ is chosen from ${\cal M}$ at random (bottom-left). An offspring is created for generation $t+1$ by recombining all chromosomes of $\alpha$ and $\beta$ and allowing their genes to mutate. The process is repeated until $M$ offspring are created. }
	\label{fig:fig02}
\end{figure}

A schematic summary of the 3CADH model is shown in Figure~\ref{fig:fig02}. 
Since mating compatibility between individuals in the 3CADH model concerns only the compatibility chromosome, we use such a condition to define reproductively isolated species. Genetic distance thresholds for the neutral ($g_{N}$), assortative ($g_{A}$) and full genome ($g_F$) can also be defined, although they have no effect in the mating dynamics and, therefore, on evolution. These thresholds can be used to group individuals according to the assortative, neutral and full set of traits and to describe correlations between the three chromosomes and their evolution through generations. A key feature of the 3CADH model is that the minimum number of loci in the $C$ chromosome necessary to induce speciation drops drastically when assortativity is taken into account. A summary of the parameters used in the 3CADH model is shown in Table~\ref{tab:tab01}.

\begin{table}
	\centering
	\begin{tabular}{cclll}
		\hline
		\multicolumn{1}{l}{} & \multicolumn{3}{c}{Description}                           & \multicolumn{1}{c}{Color}           \\ \hline
		M                    & \multicolumn{3}{l}{Number of individuals}                 &                                     \\
		$\mu$                & \multicolumn{3}{l}{Mutation probability}                         &                                     \\ \hline
		F                    & \multicolumn{2}{c}{} & Full genome                        & {\color[HTML]{228B22} Green} \\
		C                    & \multicolumn{2}{c}{} & Compatibility chromosome           & {\color[HTML]{DC143C} Crimson}      \\
		A                    & \multicolumn{2}{c}{} & Assortative chromosome           & {\color[HTML]{007FFF} Azure}        \\
		N                               & \multicolumn{2}{c}{\multirow{-4}{*}{Number of genes}}                    & Neutral chromosome           & {\color[HTML]{fcab10} Naples} \\ \hline
		& \multicolumn{2}{c}{} & Full genome $d_{F}$              &                                     \\
		& \multicolumn{2}{c}{} & Compatibility chromosome $d_{C}$ &                                     \\
		& \multicolumn{2}{c}{} & Assortative chromosome $d_{A}$ &                                     \\
		\multirow{-4}{*}{$d_{\gamma}$} & \multicolumn{2}{c}{\multirow{-4}{*}{Genetic distance~\eqref{eq:hamming}}} & Neutral chromosome $d_{N}$ &                               \\ \hline
		& \multicolumn{2}{c}{} & Full genome $g_{F}$              &                                     \\
		& \multicolumn{2}{c}{} & Compatibility chromosome $g_{C}$ &                                     \\
		& \multicolumn{2}{c}{} & Assortative chromosome $g_{A}$ &                                     \\
		\multirow{-4}{*}{$g_{\gamma}$} & \multicolumn{2}{c}{\multirow{-4}{*}{Clustering threshold}}               & Neutral chromosome $g_{N}$ &                               \\ \hline
	\end{tabular}
	\caption{Summary and description of parameters used in the models. Colors are used to differentiate between chromosomes along the simulations.}
	\label{tab:tab01}
\end{table}

\subsection*{The power of assortativity and the hitchhiking effect}

In this section we show numerical simulations of the 3CADH model for different chromosome sizes, focusing on how strict assortativity ($r=0$) facilitates speciation in a sympatric scenario. In all cases we keep the total number of genes $F=C+A+N$ fixed and change only the proportion of genes in each chromosome in the genetic architecture of the individuals. 

As the population evolves, mutations are accumulated and transmitted to offspring via reproduction.  Depending on the model parameters, the population may split into species. For the original DH model with sexual reproduction and mating restriction ($A=N=0$), species appear only if $F > 4,000$ (for a population size of $M=500$ and a mutation rate of $\mu=10^{-3}$). In our simulations we fixed $F=2,500$ to prevent speciation in the original DH models and force the process to be fully dependent on assortativity. We also fixed the population size to $M=500$, mutation rate $\mu = 10^{-3}$ and evolved the population for $T=500$ generations, which is enough to observe equilibrium in species richness in all cases studied. The initial population is homogeneous, consisting of $M$ identical individuals, with all genes set to $+1$. A common criticism to sympatric speciation modeling is the abnormal high mutation rate required for species to form. In the presented model, a lower mutation rate could still split the population into clusters (here called species), however the simulation cost would be very large \cite{baptestini2013conditions}.

\begin{figure}
	\centering
	\includegraphics[width = \textwidth]{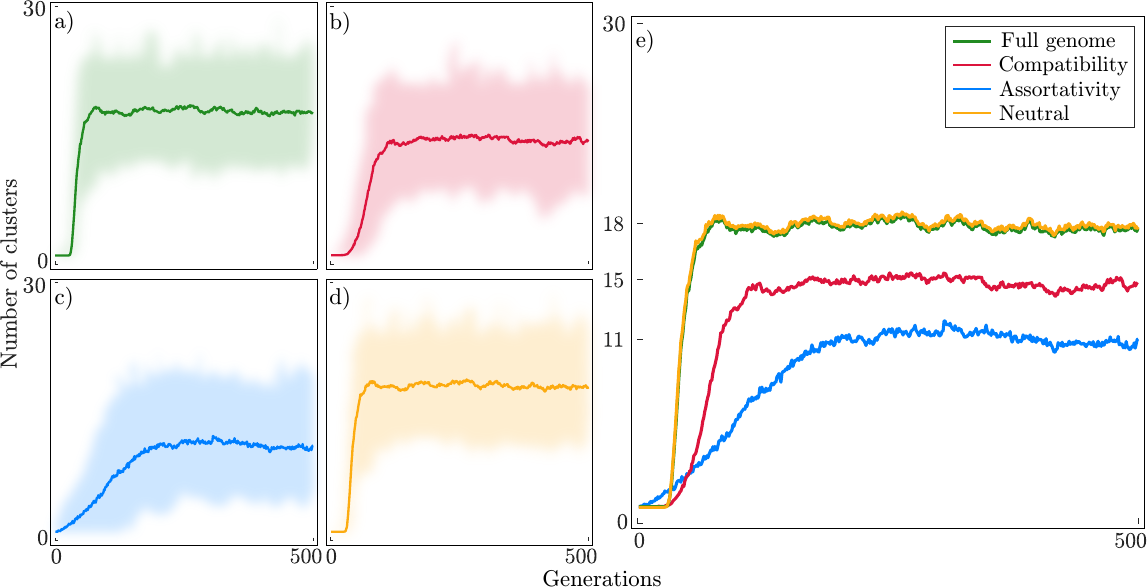}
	\caption{Evolution of the number of clusters emerged using the 3CADH model. Simulations were performed along $T = 500$ generations for $F = 2500$, $ A = 100$, $C = 100$, $r = 0$, and $\mu = 10^{-3}$. In a), b), c), and d), bold lines represent the average of $100$ executions of the model while shadows around them represent the dispersions. In e) all the average curves are put together.}
	\label{fig:fig03}
\end{figure}

The classification of individuals into clusters of similar traits can be performed for each chromosome separately and for the full genome. Classification with respect to the compatibility chromosome results in species; clusters classified according to the assortativity and neutral chromosome will be termed {\it A-clusters} and {\it N-clusters} respectively. As the population evolves, the number of clusters formed by similar individuals changes and allows us to see transitions associated to each chromosome type. Starting from a single cluster formed by the $M$, in this case 500, initially identical individuals, the average number of clusters evolves and reaches a plateau. In a single simulation, oscillations with considerable amplitude are observed, but the average over many simulations shows a smooth behavior. In Fig.~\ref{fig:fig03}, we show how the number of clusters evolve for an extreme case where assortativity and compatibility chromosome sizes are $A=C=100$, i.e., only $4\%$ of the full genome as in Fig. S1(b). Notice that the size of the compatibility chromosome is very small compared with the minimum size 4,000 needed for speciation to occur in the original DH model, showing the power of assortativity in promoting speciation in a sympatric scenario.

We used the same proportion of chromosome size to set the thresholds: $g_{N} = g_{C} = g_{F} = 0.05$. For assortativity, however, we set the threshold $g_{A} = 0.01$ so as to have the same behavior and similar scale of {\it A-clusters} as obtained by {\it C-clusters} (real species). As shown in Figs.~\ref{fig:fig03}a,~\ref{fig:fig03}b,~\ref{fig:fig03}c, and~\ref{fig:fig03}d, the average number of clusters segregated according to each chromosome converges towards a fixed number, and as a byproduct, that of the full genome also does. Note that, from Fig.~\ref{fig:fig03}e, the behavior of the number of species works as a predictor for the behavior of the full genome, even for small compatibility chromosome sizes. Therefore, the neutral genes, which make up most of the genome in these simulations, follow the behavior of the parts under selection, in a sort of hitchhiking effect. Classifying species by the neutral trait would provide results very similar to the `true' classification by reproductive isolation. Thus, neutral traits can work as proxies for reproductive isolation and, therefore, for species identification. The assortativity chromosome can also be used for species identification, although with less accuracy, as the predicted number of species depends more critically on $g_A$.

\subsection*{Species identification}

\begin{figure}
	\centering
	\includegraphics[width = \textwidth]{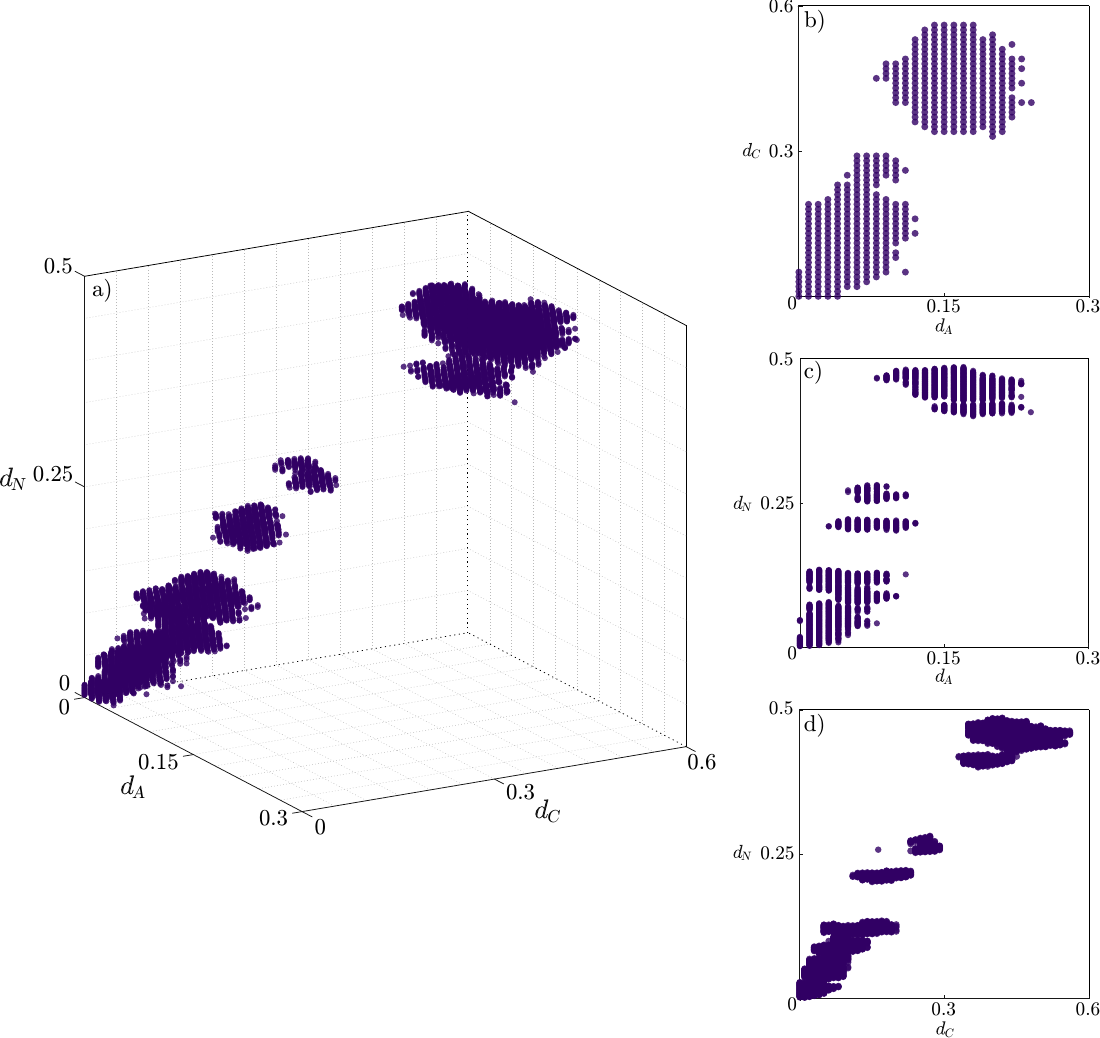}
	\caption{Genetic distance correlation between chromosomes of each individual at $t = 500$. Simulations of the 3CADH model were performed for $F = 2500$, $ A = 100$, $C = 100$, $r = 0$, and $\mu = 10^{-3}$. Circles are positioned at the genetic distance between respective chromosomes of each pair of individuals in a) a three-dimensional box and in b), c), and d) projections in two-dimensions.}
	\label{fig:fig04}
\end{figure}

Fig.~\ref{fig:fig03} suggests that genetic distances associated to all chromosomes are correlated. Indeed, once species form, reproduction is restricted to occur among members of each species, interrupting gene flow between species. As genomes of individuals belonging to different species become independent, they drift away from each other, whereas those of the same species keep mixing by reproduction, retaining some similarity. Since this argument holds for the entire genome, the similarity imposed by the mating restriction on the compatibility chromosome spreads to the assortativity and neutral chromosomes. Fig. \ref{fig:fig04} shows that genetic distances between pairs of individuals for each chromosome are actually linearly correlated. The clusters observed in this figure, that look like steam coming out of a kettle, are similar to the peaks in the histograms shown in Fig. \ref{fig:fig01}. Points close to the origin represent pairs of similar individuals, that belong to the same species. 

\begin{figure}
	\centering
	\includegraphics[width=\textwidth]{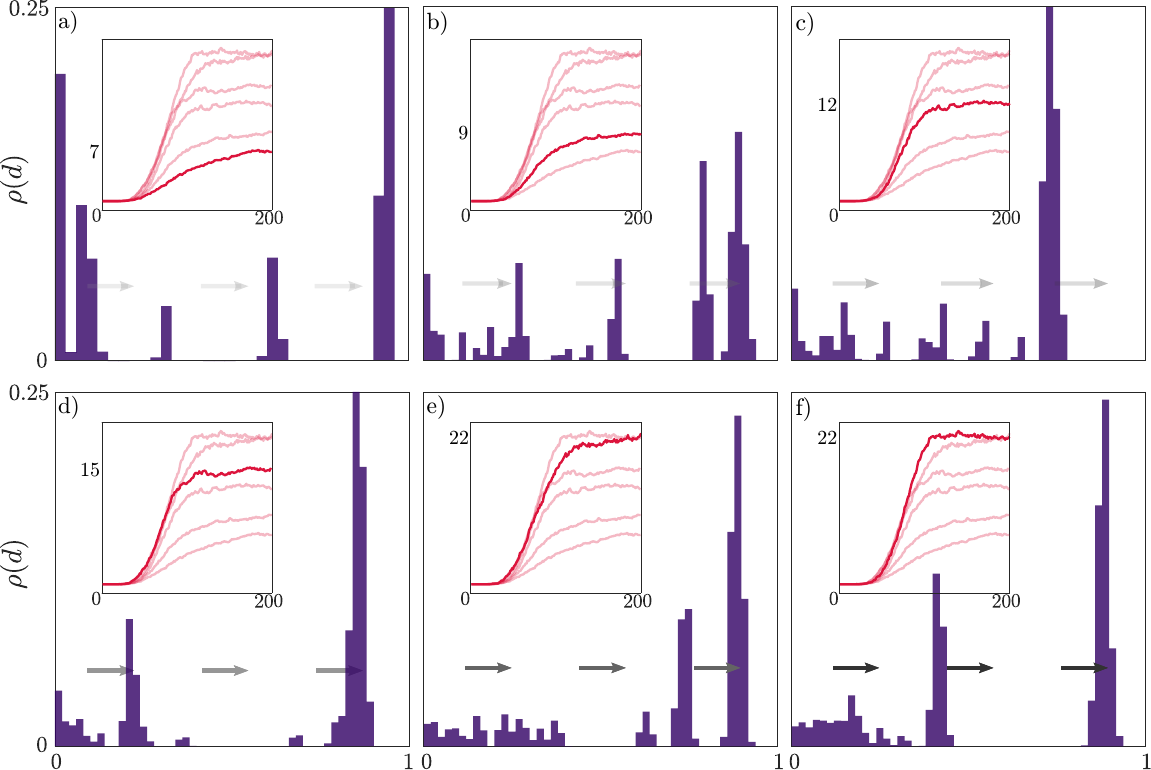}
	\caption{Genetic distance distributions in the 3CADH model at $t = 1000$ for a) $A = 10$, b) $A = 20$, c) $A = 50$, d) $A = 100$, e) $A = 1000$, and f) $A = 2400$.  Simulations were performed for $M = 500$, $F = 2500$, $r = 0$, $\mu = 10^{-3}$, and $C = 100$. In each case, embedded figures show the average number of species as a function of the time, with the bold line representing the curve for the corresponding set of parameters. The light curves show the evolution for  parameters in the other panels, for comparison. Bolder arrows indicate that the distribution drifting is faster.} 
	\label{fig:fig05}
\end{figure}

As shown previously, the use of the 3CADH model allows the emergence of species even when genome sizes are much below the threshold for speciation required by the DH model. The question arises on how large the assortativity chromosome should be for the model to preserve the speciation feature. To address this question, we show in Figure~\ref{fig:fig05} the number of species according to the 3CADH model for different sizes of the assortativity chromosome and fixed size of the compatibility for mating chromosome $C=100$. It shows that speciation occurs when the assortativity chromosome size is in a range from long enough to complement the genome without a neutral chromosome, to as small as 0.4\% of the genome. Moreover, we note a few variations in the evolution of the species as the assortativity chromosome size changes. We start by pointing at the similarities in the structure of genetic distance distributions. Similar to the single-chromosome model, there is a single peak close to 1, which arises due to the differences between species clusters, and small peaks drifting from 0 to 1, evidencing the similarities between individuals within the species. In these cases, however, the drifting of the distribution, which leads to the formation of the single peak on the right, is faster when the assortativity chromosome is longer. This response highlights the ease of forming species when mating choosiness guarantees that sexual partners will be most similar to focal individuals. From the cluster transitions, we also notice a variation in the number of species dependent on the assortativity chromosome size. When the assortativity chromosome size surpasses this threshold, the number of species converges to a fixed number, as opposed to when the chromosome size is below, where the number of species is dependent on it.

\begin{figure}
	\centering
	\includegraphics[width = 0.6\textwidth]{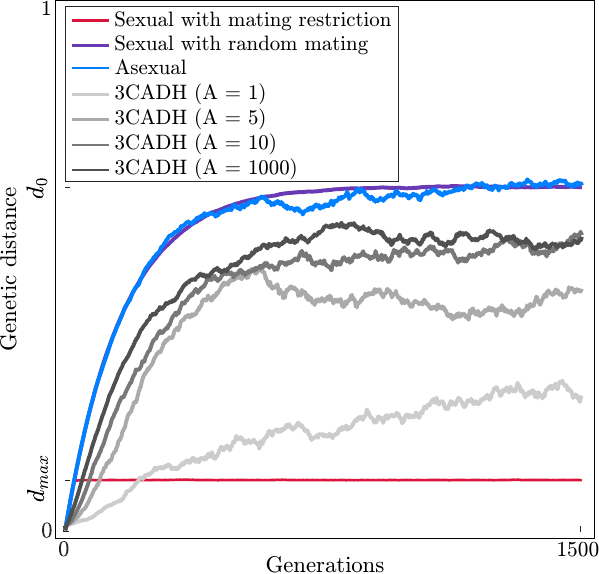}
	\caption{Average genetic distance for the asexual, sexual with random mating, sexual with mating restriction, and 3CADH cases. Each case shows an average of 50 executions along $T = 1500$ generations for $M = 500$ and $F = 2500$.}
	\label{fig:fig06}
\end{figure}

All the effects previously described can be addressed by understanding the linkage between chromosomes in the 3CADH process. If the assortativity chromosome were much larger than the others, the process would be almost identical to asexual reproduction. Despite finding a first compatibility subset, the second step of finding the most similar partner with the assortativity chromosome
would imply a high level of genetic similarity between the mating pair. The effect of the assortativity chromosome driving speciation is shown in Figure~\ref{fig:fig06}, where the system exhibits a stable behavior even in the case where only one gene determines assortativity. The averaged genetic distance of the DH model converges to the threshold value $d_{max}$ due that the genome size is not long enough to generate species. The asexual and homogeneous models show the convergence of the averaged genetic distance to the mean value $d_{0}$. Whereas, for different sizes of assortativity chromosome, the convergence of the averaged genetic distance ranges from $d_{max}$ to $d_{0}$. 

\subsection*{The trouble with assortativity}

In the previous sections we have shown results for strict assortativity only: among all compatible individuals (as dictated by the compatibility chromosome C) only the most similar to the mating individual (with respect to A) have a chance of reproducing with it. Those differing by a single extra allele would never be chosen. Here we relax this condition letting the choosiness parameter $r$ be positive. 

To have an idea of the effects of $r$ we note that for $A=100$ and $r=0.02A$, we include as potential mating partners individuals differing by 1 or 2 extra alleles beyond the maximum similarity at that moment. Note that the number of extra individuals included in the set increases rapidly as $r$ increases. Therefore, we expect assortativity to stop working as a powerful selection mechanism quite soon. 

\begin{figure}
	\centering
	\includegraphics{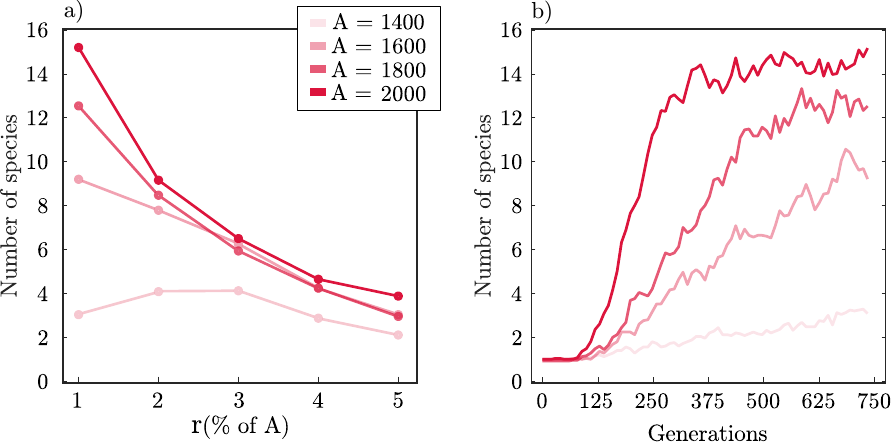}
	\caption{Effects of choosiness on the emergence of species. Simulations were performed for $M = 500$, $F = 2500$, $\mu = 10^{-3}$, and $C = 100$. Both figures picture results from averaging 25 executions for the respective parameters. In a), each dot represents the number of species after $750$ generations for different values of $A$ and $r$. In b), each line describes the generational evolution of the number of species for different values of $A$ and fixed $r = 0.01A$.}
	\label{fig:fig07}
\end{figure}

Figure \ref{fig:fig07} shows the effect of $r$ on the number of species for different sizes of the assortativity chromosome. We observe that indeed $r$ leads to a rapid decrease of the number of species, especially for small values of $A$. Besides, it also delays speciation and slows down  equilibration.

\section*{Discussion}

We have studied the effects of assortativity in the emergence of species in a sympatric population of hermaphrodites individuals that reproduce sexually. Our work is based on the model proposed by Higgs and Derrida (DH) \cite{higgs1991stochastic}, where individuals are represented by biallelic genomes and reproduction requires the genomes of the mating pair to be sufficiently similar (genetic compatibility). Speciation in this scenario only happens if the genome size is sufficiently large (of the order of 4000 genes for the parameters used in our work). Here, we generalize the representation of the individuals and the way mating partners are selected. Individuals are represented by three chromosomes: a compatibility $C$, an assortativity $A$, and a neutral $N$ chromosome. As in the DH model, mating is allowed only between compatible individuals, with Hamming distance between their $C$ chromosomes smaller than a threshold value $G$. However, among the compatibles, the one with maximum genetic similarity with respect to the $A$ chromosome is chosen.  

We have shown that, in our model, speciation occurs even when genome size is small, much below the threshold required for the original DH model ~\cite{de2017speciation}. This drastic change in genome size requirement for speciation is a consequence of assortativity and the corresponding decrease in the gene flow it promotes. Then, under strict conditions, assortativity is shown to be sufficient for the emergence and coexistence of species. This differs from previous results~\cite{bagnoli2005model}, where assortativity alone promotes the emergence of species, but these vanish in the absence of competition. We have also shown that species, which are classified according to reproductive isolation imposed by the compatibility chromosome, can also be identified by comparing the $N$ chromosome, which works as a proxy for reproductive isolation.

The dynamics of species formation in our model can be understood as follows: in the beginning of the simulation all chromosomes are identical and, therefore, all individuals are compatible. Mutations introduce small differences that, although are not enough to create incompatibilities during the first generations, produce different versions of the assortativiy chromosome. Individuals with a unique $A$ chromosome never get selected for reproduction, and although these individuals have the chance of mating at least once, the rarity of their $A$ chromosome will culminate with its disappearance. Groups of individuals with identical $A$'s, on the other hand, will only mate among themselves, as they select the most similar possible partner. This creates clusters of individuals with very similar $A$ chromosome. The key effect of this process is to create small isolated groups that, although not reproductively isolated from rest of the population, will only mate with others from the same group, purging new mutations in chromosome $A$. After the formation of such isolated groups, further mutations and drift lead the $C$ chromosome to differentiate among groups, leading eventually to the formation of species. As shown in Figure
S2, one single locus in the $A$ chromosome guarantees the emergence of genetic variation necessary for diversification -- at least in the case of strong assortativity, as imposed by the model.


Our results also reveal that, despite each chromosome and each locus being independent, there is genetic linkage disequilibrium (LD) in all the different chromosomes. The mechanism of assortativity leads indirectly to such correlations via species formation. As a consequence both the compatibility and the neutral chromosomes can identify almost exactly the same species. Therefore, although the neutral chromosome does not participate in the dynamics of reproduction, it suffers the dynamic consequences that occur in the other chromosomes and mirrors the arrangement they went through. Assortative mating has been proposed as a mechanism that can lead to speciation, particularly in  sympatric scenarios \cite{smith1966sympatric, Felsenstein_1981, Dieckmann_1999, fry2003, Bolnick_2004}. For assortative mating to drive speciation, theoretical models often require disruptive selection, such as those proposed by Dieckmann and Doebeli \cite{Dieckmann_1999} or scenarios where different fixed alleles exist in each subpopulation \cite{Felsenstein_1981, butlin2021homage}. One of the most influential works in this area is Felsenstein's 1981 paper, where he proposed genetic forces involved in speciation that incorporated assortative mating. He developed two different models using three gene loci, where the biallelic loci $B$ and $C$ were under selection in two different subpopulations connected by migration. The $BC$ haplotype was favored in one subpopulation, while the $bc$ was favored in the other.

In the one-allele model, Felsenstein predicted that the expression of a third gene locus ($A$) coding for strong assortativity in both subpopulations would reduce gene flow and induce speciation. In the framework of our model, this would be similar to impose a spatial mating neighborhood, where individuals can find their mates. In that case gene flow is reduced by this spatial mating restriction, without the need for an assortativity chromosome, as explored in \cite{de2009global, martins2013, de2017speciation, costa2019, marquitti2020}. In the other Felsenstein's model, the two-allele model, two (or more) alleles in the third locus define the mating preference of individuals. Carriers of the allele $A$ ($a$) prefer to mate with those also carrying $A$ ($a$). If $A$ is preferred in the habitat where the $BC$ haplotype is being favored, and $a$ is preferred in the habitat with $bc$ haplotypes, then a linkage disequilibrium (LD) emerges. However, due to genetic recombination dissolving the LD, Felsenstein predicted that this model was unlikely to lead to speciation \cite{kopp2018mechanisms, butlin2021homage, Felsenstein_1981}.

The model explored in this work resembles the two-allele model, but with more alleles. We show, in contrast to Felsenstein's predictions, that assortative mating induces speciation in this sympatric scenario. Additionally, we do not require disruptive selection or differential selection by habitat for species to form. However, we do impose a genetic threshold in the compatibility chromosome, which defines the pool of compatible individuals and is also used for species definition. Assortative mating acts as the initial spark for species formation, reducing gene flow between dissimilar individuals and allowing speciation to occur. We observed LD among different chromosomes identifying almost the same species. Since during sexual mating offspring inherit each locus independently from their parents, we see that sexual reproduction is not breaking genetic associations that lead to speciation.  We are aware of the lack of evidence for the two-allele model. However, as Servedio and Noor (2003) have pointed out \cite{servedio2003role}, it is very difficult to distinguish whether the process by which a group of species evolved was a one-allele or two-allele scenario. 

We expect that the gene flow interruption induced by assortativity may persist to some degree if this mating mechanism disappears due to biological cost or selection. To further investigate the effects of choosiness in reducing gene flow and creating initial clusters of individuals, future studies could relax the strong assortativity we imposed by using a probability distribution or by creating pools of possible partners \cite{caetano2020sympatric}. Additionally, decreasing assortativity after species have formed could help shed light on the significance of choosiness for species maintenance.

Sympatric speciation is a controversial topic \cite{coyne2000little, bolnick2007sympatric, fitzpatrick2008if} due to the difficulty in gathering information and fossilized evidence, as well as the challenges of manipulating complex life forms in laboratory settings. However, as science advances, it also creates the necessary tools for solving old problems and analysing classic questions with less limitations and genomics is an allied for this open problem in the coming years \cite{foote2018sympatric, richards2019searching}. Despite these limitations, we recognize the importance of theoretical approaches in helping to understand and potentially recognize this process in nature. \\

\noindent{\textbf{Data Availability Statement:}} No data was used in this work.

\section*{Acknowledgments}

\noindent This work was partly supported by FAPESP grant 2021/14335-0 and CNPq grant 301082/2019‐7 (MAMA), by the Coordenação de Aperfeiçoamento de Pessoal de Nível Superior -- Brasil (CAPES) -- Finance Code 001 (FMDM), and by FAPESP grant 2021/04251-4 (JUFL).

\setstretch{1.5}

\end{document}